\documentclass{article}
\usepackage[utf8]{inputenc}
\usepackage{amsmath,amssymb,pxfonts,color,ulem}
\usepackage{booktabs}
\usepackage{pdflscape}
\usepackage{algpseudocode}
\usepackage{algorithm}
\usepackage{hyperref}
\newcommand{\nothing}[1]{}
\usepackage{graphicx}

\newtheorem{rem}{Remark}
\newtheorem{lemma}{Lemma}
\newtheorem{ex}{Example}
\newtheorem{thm}{Theorem}

\title{Liquidity provision  of utility indifference type
in decentralized exchanges}
\author{Masaaki Fukasawa\thanks{The University of Osaka, Japan. Email: fukasawa@sigmath.es.osaka-u.ac.jp}, Basile Maire\thanks{Quantena, Switzerland.}, and Marcus Wunsch\thanks{ZHAW School of Management and Law, Switzerland}.}

\date{}
\begin{document}

\maketitle

\begin{abstract}
\noindent
We present a mathematical formulation of liquidity provision in decentralized exchanges. 
We focus on constant function market makers of utility indifference type, which include constant product market makers with concentrated liquidity as a special case.
First, we examine no-arbitrage conditions for a liquidity pool and compute an optimal arbitrage strategy when there is an external liquid market.
Second, we show that liquidity provision suffers from impermanent loss unless a transaction fee is levied under the general framework with concentrated liquidity. 
Third, we establish the well-definedness of arbitrage-free reserve processes of a liquidity pool in continuous-time and show that there is no loss-versus-rebalancing under a nonzero fee if the external market price is continuous.
We then argue that liquidity provision by multiple liquidity providers can be understood as liquidity provision by a representative liquidity provider, meaning that the analysis boils down to that for a single liquidity provider. 
Last, but not least, we give an answer to the fundamental question in which sense the very construction of constant function market makers with concentrated liquidity in the popular platform Uniswap v3 is optimal. 
\end{abstract}
\bigskip
\noindent{\it Keywords: 
Decentralized finance, automated market makers, arbitrage, impermanent Loss, divergence loss}
\\

\noindent{\it JEL Classification Codes: D47, D53, C02}

\newpage
\section{Introduction}
\subsection{Decentralized Exchange}
Decentralized exchanges (DEXs) represent a significant evolution in the realm of digital finance, harnessing blockchain technology to facilitate peer-to-peer trading without the need for centralized intermediaries. Unlike traditional centralized exchanges, which control user funds and facilitate transactions via an intermediary, DEXs operate on a decentralized network of nodes, ensuring greater transparency, security, and autonomy for users.

At the core of DEXs lies the use of smart contracts,
self-executing contracts with the terms of the agreement directly written into code. These smart contracts enable automated and trustless transactions, reducing the risk of hacks and manipulation associated with centralized platforms. Moreover, DEXs offer users full control over their assets, enhancing privacy and reducing dependence on any single point of failure.

One of the prominent models utilized by DEXs is the Automated Market Maker (AMM). AMMs rely on Liquidity Pools (LPs) and algorithms to facilitate trading without the need for a traditional order book. Instead of matching buyers and sellers, AMMs use a deterministic pricing algorithm based on the ratio of assets in the liquidity pool. This mathematical approach is often exemplified by the constant product formula $xy = k$, where $x$ and $y$ are the quantities of two assets and $k$ is a constant, ensuring that the product of the assets' quantities remains invariant.

The mathematical analysis of AMMs involves studying the properties of these pricing algorithms, their impact on liquidity, slippage, and impermanent loss. Impermanent loss, for example, is a phenomenon where liquidity providers may experience a reduction in the value of their assets compared to simply holding them. This occurs due to the divergence in the prices of the pooled assets, necessitating a detailed understanding of the underlying mathematical principles to mitigate risks.

\subsection{Constant Function Market Maker}
In DEX such as Uniswap v3, an LP consists of a pair of digital assets that are deposited as reserves by liquidity providers. 
The AMM associated with this pool is a smart contract that executes orders from liquidity takers. 
A general class of AMMs are so-called Constant Function Market Makers (CFMM), where liquidity takers swap $\xi$ units of the first currency for $\eta$ units of the second currency in accordance with the equation
\begin{equation}\label{CFMM}
   \varphi(x,y, \xi,\eta) = \varphi(x,y,0,0),
\end{equation}
where $\varphi$, called the \textit{trading function},
is a function which is increasing with respect to each of its arguments,
and $x\geq 0$ and $y \geq 0$ are, respectively, the reserve amounts of the first and second currency in the LP at the time.
When $\xi >0$ (resp. $\xi < 0$), this means a liquidity taker pays $\xi$  units of the first (resp. $\eta >0 $ units of the second) currency to the LP to receive $-\eta > 0$  units of the second (resp. $-\xi > 0$ units of the first) currency from the LP. 
In case the LP levies fees in a Uniswap v3-type architecture, letting $\tau \in (0,1)$ denote a fee coefficient,
when $\xi >0$ (resp. $\xi < 0$), $(1-\tau)\xi$ units of the first (resp. $(1-\tau)\eta$ units of the second) currency are added to the LP and the remained $\tau \xi$ units (resp. $\tau \eta$ units) are pooled in a separate fee collecting account. (This is in contrast to Uniswap v2, where fees are collected within the same pool.)

\subsection{Contribution of the paper}
Many papers that have appeared recently to analyze AMMs, assume the internal price, that is, the infinitesimal exchange ratio of the LP, coincides with the price in an external liquid market even for an LP with nonzero transaction fee, cf.~\cite{AEC,CDM,CDM2,CSSS,E, EAC, FMW1, MMRZ,MMR, M, TW}. 
This is a relevant no-arbitrage assumption under zero fee; however under nonzero fee, the internal price has a finite total variation and so, it does not coincide with the external price, as illustrated by \cite{FMW2}. See also \cite{NTYY,LTW}. We extend the preceding works to a rigorous treatment of AMMs with concentrated liquidity such as Uniswap v3. 
We provide a general mathematical framework and show that the impermanent loss can be super-hedged by a model-free rebalancing strategy in the external market irrespectively of the size of transaction fee if the external price is continuous.

In Section~2, we give a mathematical formulation of AMMs of utility indifference type, which is a general class of Constant Function Market Makers including Constant Product Market Makers with Concentrated Liquidity as an example (Remark~\ref{remXi}).
In Section~3, as a preliminary analysis, we examine the no-arbitrage conditions and compute the optimal arbitrage strategy for both types of condition violations.
In Section~4, we consider cases with zero transaction fee.
Theorem~\ref{thm1} shows that the value of an LP can be represented as a Legendre transform, extending \cite{AEC,FMW1} to cases with concentrated liquidity, which implies that the value is a concave function of the external price, resulting in impermanent loss\footnote{The shortfall of liquidity provision versus buy-and-hold, cf.~\cite{M}} (Remark~\ref{IL1}).
Theorem~2 gives a representation of {\it loss-versus-rebalancing} extending \cite{MMRZ,FMW1,CDM2},  to cases with concentrated liquidity when the external price is a continuous semimartingale.
In Section~5, we extend the analysis for Uniswap v2 type architecture of \cite{FMW2} that incorporates
transaction fees, to a Uniswap v3-type architecture in which transaction fees are collected outside the LP.
By Theorem~\ref{thm:main}, we establishes the well-definedness of arbitrage-free reserve processes of an LP in continuous-time, and show that there is no loss-versus-rebalancing under nonzero fee if the external market price is continuous.
In Section~6, we argue that an LP with multiple liquidity providers, each having a bespoke liquidity provision range as in Uniswap v3, can be represented by a single, representative liquidity provider, reducing the analysis to that for a single liquidity provider as discussed in the preceding sections.

\section{Trading function of utility indifference type}
Here, we give a mathematical model of an AMM.
We focus on the trading function of the form
\begin{equation}\label{varphi}
\begin{split}
     \varphi(x,y,\xi, \eta)
    =& u_\ast(x + (1-\tau H(\xi)) \xi,  y+ (1-\tau H(\eta))\eta) 
\end{split}
\end{equation}
where $u_\ast(x,y)= u(x_\ast + x, y_\ast + y)$,
$u:(0,\infty)\times (0,\infty) \to \mathbb{R}$ is a strictly concave 
three times continuously differentiable function which is
increasing in both of the arguments with
\begin{equation*}
\lim_{x\downarrow 0} u(x,y) =     \lim_{y\downarrow 0} u(x,y), \ \ 
\lim_{x\uparrow \infty} u(x,y) =     \lim_{y\uparrow \infty} u(x,y),
\end{equation*}
$x_\ast \in [0,\infty)$, $y_\ast \in [0,\infty)$ and
$\tau \in [0,1)$ are constants, and $H$ is the Heaviside function; $H(z) = 1$ for $z \geq 0$ and $=0$ for $z < 0$.  
The parameters $(x_\ast, y_\ast)$ and $\tau$ control, respectively, the range of liquidity provision  and the size of transaction fee as seen more clearly later.

The simplest example of $u$ is $u(x,y) = xy$, which corresponds to the Constant Product Market Maker.
The Cobb-Douglas utility function
$u(x,y) = x^\alpha y^{1-\alpha}$, $\alpha \in (0,1)$, corresponds to the class of Geometric Mean Market Makers.
 Considering the case of $\tau=0$ in \eqref{varphi}, that is,
 \begin{equation*}
     \varphi(x,y,\xi,\eta) = u_\ast(x+\xi, y + \eta),
 \end{equation*}
the rule \eqref{CFMM} is understood as a utility indifference principle in market making,
with the utility function $u_\ast$.

When \eqref{CFMM} is met for a trading function  
$\varphi$ of the form \eqref{varphi}, we have 
\begin{equation}\label{CFMM2}
    u_\ast(x, y) 
    = \begin{cases}
          u_\ast(x + (1-\tau)\xi, y+\eta) & \xi >0, \\
 u_\ast(x + \xi, y+(1-\tau)\eta) & \xi < 0
    \end{cases}
    \end{equation}
by the definition of the Heaviside function.
We consider in this paper such an LP that its reserves $(x,y)$ are updated to 
\begin{equation*}
    (x+ (1-\tau H(\xi)) \xi,  y+ (1-\tau H(\eta))\eta)
    = 
     \begin{cases}
          (x + (1-\tau)\xi, y+\eta) & \xi >0, \\
 (x + \xi, y+(1-\tau)\eta) & \xi < 0
    \end{cases}
\end{equation*}
after executing a swap order $(\xi, \eta)$ by a liquidity taker, 
so that the value of $u_\ast$ is kept unchanged.
In particular, the AMM only accepts orders subject to 
\begin{equation}\label{CFMM3}
x+ (1-\tau H(\xi)) \xi\geq 0, \ \ 
y+ (1-\tau H(\eta))\eta \geq 0.
\end{equation}

The constant $\tau$ controls the size of the transaction fees.
There is a separate fee collection account for the LP,  where 
the cumulative fees $(x^f,y^f)$ are updated to
$$(x^f+  \tau H(\xi) \xi,  y^f+ \tau H(\eta)\eta)$$
by swap  $(\xi,\eta)$.
The wealth of the liquidity provider consists of
$x + x_f$ units of the first and
$y+ y_f$ units of the second currencies
when the LP's reserves and cumulative fees are, respectively, $(x,y)$ and $(x_f,y_f)$.

By the implicit function theorem, for any $(x_0,y_0) \in (0,\infty)\times (0,\infty)$, there exists a twice differentiable decreasing bijection $f:(0,\infty) \to (0,\infty)$ such that
\begin{equation*}
    u(x,f(x)) = u(x_0,y_0), \ \ x \in (0,\infty).
\end{equation*}
By the strict concavity of $u$, $f$ is strictly convex.
Let $(x,y)$ be the current LP  reserves and 
$f$ be the implicit function with respect to
$(x_0,y_0) = (x_\ast + x, y_\ast + y)$. 
Then, \eqref{CFMM2} and \eqref{CFMM3} are respectively written in terms of $f$ as
\begin{equation*}
    \eta = \eta(\xi):= \begin{cases}
      f(x_0 + (1-\tau)\xi) - f(x_0)  & \xi > 0, \\
      (f(x_0 + \xi) - f(x_0))/(1-\tau) & \xi < 0
    \end{cases}
\end{equation*}
and
\begin{equation*}
    \begin{cases}
        f(x_0 + (1-\tau)\xi) \geq y_\ast  & \xi > 0, \\
       x_0 + \xi \geq x_\ast & \xi < 0.
    \end{cases}
\end{equation*}
\begin{ex}
\upshape
If $u(x,y) = xy$, then
$u_\ast(x,y) = (x+x_\ast)(y+y_\ast)$ and
$f(x) = L/x$, where $L = u(x_0,y_0)=x_0y_0$. 
Other examples can be found in \cite{BF}.
\end{ex}

Let $(X_t,Y_t)$ denote the reserves at time $t \geq 0$ in the LP.
We assume  $X_0 \geq 0$, $Y_0 \geq 0$ and
$(x_0,y_0) := (x_\ast + X_0,y_\ast + Y_0) \in (0,\infty)\times (0,\infty)$.
Since $u_\ast(X_t,Y_t)$ is kept constant by the AMM algorithm, we have 
$Y_t = f_\ast(X_t) $ for all $t$, where $f_\ast(x) = f(x_\ast + x) -y_\ast$ and 
$f$ is the implicit function of $u$ with respect to
$(x_0,y_0)$.
Let 
\begin{equation*}
  \Xi  = (- x_\ast,\infty) \cap [0,x_\dagger], \ \   x_\dagger = f_\ast^{-1}(0+) := \lim_{y \downarrow 0} f_\ast^{-1}(y)
\end{equation*}
and 
\begin{equation}\label{BA}
\begin{split}
s(x)  &= -f_\ast^\prime(x),\\
      a(x) & = - \lim_{\xi\uparrow 0} \frac{\eta(\xi)}{\xi} = \frac{1}{1-\tau}s(x), 
      \\ b(x)& = -\lim_{\xi \downarrow 0}  \frac{\eta(\xi)}{\xi} = (1-\tau)s(x)
\end{split}
\end{equation}
for $x \in \Xi$. 
Then, $A_t := a(X_t)$ and $B_t := b(X_t)$ are infinitesimal exchange ratios in the LP at time $t$ which we call the ask and bid prices of the first currency respectively.
Note that $s: \Xi \to s(\Xi)$ is a decreasing bijection by the strict convexity of $f$.
\begin{rem}\label{remXi}\upshape 
    If $(x_\ast,y_\ast) = (0,0)$, then $\Xi = (0,\infty)$.
    In this case,
    since $f(0+) = \infty$ and $f(\infty-) = 0$, we have
    $f^\prime(0+) = -\infty$ and $f^\prime(\infty-) = 0$, hence
    $s(\Xi) = (0,\infty)$.
    When $x_\ast > 0$, then $0 \in \Xi$. When $y_\ast > 0$, then $x_\dagger = f^{-1}(y_\ast)\in \Xi$.
    As seen in the next section,
    positive values of $x_\ast$ and $y_\ast$ realize a concentrated liquidity provision on the interval $s(\Xi) =
    [s(x_\dagger),s(0)] =:[S_L,S_U]$.
    When $u(x,y) = xy$, then $f_\ast(x) = L^2/(x+x_\ast) - y_\ast$, where
    $L^2 = (X_0 + x_\ast)(Y_0 + y_\ast)$, and so
    \begin{equation*}
x_\dagger = f_\ast^{-1}(0) = L^2/y_\ast - x_\ast,\ \          S_L= s(x_\dagger) = 
\frac{y_\ast^2}{L^2}, \ \ S_U = s(0) = \frac{L^2}{x_\ast^2}
    \end{equation*}
    and
    \begin{equation*}
        u_\ast(x,y) = u(x+x_\ast,y+y_\ast) = \left(x + \frac{L}{\sqrt{S_U}}\right)
        \left(y + L\sqrt{S_L}\right).
    \end{equation*}
\end{rem}

We assume that there are external liquid exchanges\footnote{These could be, for instance, Centralized Exchanges such as Binance or Kraken, or other DEX.} for the currency pair, and let $S^\ast = \{S^\ast_t\}$ denote the external price (exchange ratio) process. More precisely, $S^\ast_t$ is the unique price for one unit of the first currency in terms of the second in the external markets at time $t$.  
We say the LP is free of arbitrage at time $t$ if
\begin{equation}\label{NAR}
    A_t \geq S^\ast_t 
\end{equation}
is met whenever $X_t >0$ and 
\begin{equation}\label{NAL}
    B_t \leq S^\ast_t 
\end{equation}
is met whenever $Y_t>0$. 
As we see later,
there is an arbitrage opportunity when \eqref{NAR} or \eqref{NAL} is violated.
The wealth of the liquidity provider at time $t$ is evaluated in terms of the second currency as
\begin{equation}\label{vt}
 V_t := Y_t + Y^f_t  + (X_t + X^f_t)S^\ast_t,
\end{equation}
where $(X^f_t,Y^f_t)$ are the cumulative fee earnings at time $t \geq 0$.
Naturally, we assume $(X^f_0,Y^f_0) = (0,0)$ in the sequel.

\section{Optimal arbitrage}
In this section,
 we observe that if the LP is not free of arbitrage at time $t$ in the sense of \eqref{NAR} and \eqref{NAL}, there is indeed an arbitrage opportunity, and that after an optimal arbitrage trade exploiting it, the LP becomes free of arbitrage.

If \eqref{NAR} is violated while $X_t > 0$, there is an arbitrage opportunity.
To see this, suppose  $X_t >0$ and $A_t < S^\ast_t$. 
Since
\begin{equation*}
      \frac{f_\ast(X_t + \xi) - f_\ast(X_t)}{\xi} 
    \uparrow  f_\ast^\prime(X_t) = - (1-\tau)A_t
\end{equation*}
as $\xi \uparrow 0$ by the convexity of $f_\ast$,
we have
\begin{equation*}
     - \frac{f_\ast(X_t + \xi) - f_\ast(X_t)}{1-\tau} -\xi S^\ast_t 
     = -\frac{\xi}{1-\tau} \left(
         \frac{f_\ast(X_t + \xi) - f_\ast(X_t)}{\xi} 
       + (1-\tau)S^\ast_t \right)
     > 0
\end{equation*}
for sufficiently small $-\xi  > 0$.
The left hand side is the profit-and-loss in
buying $-\xi$ units of the first asset in the LP and
selling them in the external market.
This means a swap $(\xi,\eta)$ with such $\xi$ at the LP
makes an arbitrage profit.

The optimal size of $\xi$ is obtained by solving the first order condition
\begin{equation*}
  0 = \frac{\mathrm{d}}{\mathrm{d}\xi}
  \left(
   - \frac{f_\ast(X_t + \xi) - f_\ast(X_t)}{1-\tau} -\xi S^\ast_t 
 \right)= \frac{1}{1-\tau} s(X_t + \xi)- S^\ast_t.
    \end{equation*}
    Therefore, if $(1-\tau)S^\ast_t \in s(\Xi)$, 
    we expect an order $(\xi^\ast,\eta^\ast)$ from the arbitrageur after which the ask price is updated 
    from
    \begin{equation*}
        A_t = \frac{1}{1-\tau}s(X_t)
    \end{equation*}
    to
    \begin{equation*}
        \frac{1}{1-\tau}s(X_t + \xi^\ast) = S^\ast_t.
    \end{equation*}
When $x^\ast > 0$, there is a possibility that
$S^\ast_t > s(0)/(1-\tau) = a(0)$ since $0 \in \Xi$ (see Remark~\ref{remXi}).
In this case, $\xi^\ast :=  - X_t < 0$ is the optimal arbitrage order and the LP reserves is updated 
from $(X_t,Y_t)$ 
to $(X_t+ \xi^\ast,Y_t + \eta^\ast) = (0,f_\ast(0))$.  
Although the updated ask price, $a(0)$, is still below $S^\ast_t$,
no more swap $(\xi,\eta)$ with $\xi < 0$ is allowed, since the LP reserve of the first currency is now zero.

If \eqref{NAL} is violated while $Y_t > 0$, there is also an arbitrage opportunity.
To see this, suppose  $Y_t >0$ and $B_t > S^\ast_t$. 
Since
\begin{equation*}
      \frac{f_\ast(X_t + (1-\tau)\xi) - f_\ast(X_t)}{(1-\tau)\xi} 
    \downarrow  f_\ast^\prime(X_t) = -\frac{1}{1-\tau}B_t
\end{equation*}
as $\xi \downarrow 0$ by the convexity of $f_\ast$,
we have
\begin{equation*}
\begin{split}
      &-( f_\ast(X_t + (1-\tau)\xi) - f_\ast(X_t)) -\xi S^\ast_t 
     \\ & = \xi(1-\tau) \left( -
         \frac{f_\ast(X_t + (1-\tau)\xi) - f_\ast(X_t)}{(1-\tau)\xi} 
       -  \frac{1}{1-\tau}S^\ast_t \right)
     > 0
\end{split}
\end{equation*}
for sufficiently small $\xi  > 0$.
The left hand side is the profit-and-loss in
buying $\xi$ units of the first asset in the external market and
selling them in the LP.
This means a swap $(\xi,\eta)$ with such $\xi$ at the LP
makes an arbitrage profit.
The optimal size of $\xi$ is obtained by solving the first order condition
\begin{equation*}
  0 = \frac{\mathrm{d}}{\mathrm{d}\xi}
  \left(
   - (f_\ast(X_t + (1-\tau)\xi) - f_\ast(X_t)) -\xi S^\ast_t 
 \right)= (1-\tau) s(X_t + \xi)- S^\ast_t.
    \end{equation*}
    Therefore, if $S^\ast_t/(1-\tau) \in s(\Xi)$, 
    we expect an order $(\xi^\ast,\eta^\ast)$ from the arbitrageur after which the bid price is updated
    from
    \begin{equation*}
        B_t = (1-\tau) s(X_t)
    \end{equation*}
    to
    \begin{equation*}
        (1-\tau)s(X_t + \xi^\ast) = S^\ast_t.
    \end{equation*}
When $y^\ast > 0$, there is a possibility that
$S^\ast_t < (1-\tau)s(x_\dagger) = b(x_\dagger)$ since $x_\dagger  = f^{-1}(y_\ast)\in \Xi$ (see Remark~\ref{remXi}).
In this case, $\xi^\ast :=  x_\dagger- X_t > 0$ is the optimal arbitrage order and the LP reserves are updated to $(X_t+ \xi^\ast,Y_t + \eta^\ast) = (x_\dagger,0)$.   
Although the updated bid price, $b(x_\dagger)$, is still above $S^\ast_t$,
no more swap $(\xi,\eta)$ with $\eta < 0$ is allowed, since the LP reserve of the second currency is now zero.

\section{Impermanent loss under zero fee}
We assume $\tau = 0$ throughout this section.
In this case, there is no fee collection and so,
$X^f = Y^f = 0$. The wealth process of the liquidity provider is 
then $V = Y + XS^\ast$ by \eqref{vt}.
Let $l = \inf s(\Xi)$, $r = \sup s(\Xi)$ and define
\begin{equation}\label{eqS}
    S_t = (S^\ast_t \vee l) \wedge r = l + (S^\ast_t-l)_+ - (S^\ast_t-r)_+.
\end{equation}
When $r = \infty$, $(S^\ast_t -r)_+$ is interpreted as $0$.
Note that $l = s(x_\dagger-)$ and $r = s(0+)$,
and that if $S^\ast_t \in s(\Xi)$, then $S^\ast_t = S_t$.
\begin{lemma}\label{lem1}
    If 
the LP is free of arbitrage at time $t$, then
$S_t = s(X_t)$.
\end{lemma}
The proof is given in Section~\ref{prlem1}.

\begin{thm}
\label{thm1}
If $S^\ast_t \in s(\Xi)$ and 
the LP is free of arbitrage at time $t$, then
    $V_t = v(S^\ast_t)$ and $X_t = v^\prime(S^\ast_t)$, where
    \begin{equation}\label{vp}
        v(p) = \inf_{x \in \Xi}\{xp + f_\ast(x)\}.
    \end{equation}
\end{thm}
The proof is given in Section~\ref{prthm1}.

\begin{rem}\label{IL1}
\upshape
Note that
the function $v$ defined by \eqref{vp} is concave 
(because it is the infimum of linear functions)
and so
$v(S^\ast_t) \leq v(S^\ast_0) + v^\prime(S^\ast_0)(S^\ast_t-S^\ast_0)$
if $S^\ast_t, S^\ast_0 \in s(\Xi)$.
Since $v^\prime(S^\ast_0) = X_0$, this inequality implies the impermanent loss
\begin{equation*}
Y_0 + X_0S^\ast_t- V_t \geq 
Y_0 + X_0S^\ast_t 
-V_0 - X_0(S^\ast_t-S^\ast_0) = 0,
\end{equation*}
meaning that the wealth $V_t$ by the liquidity provision is always inferior to
the buy-and-hold wealth
$Y_0 + X_0S^\ast_t$.
\end{rem}

\begin{thm}\label{thmn}
    If the LP is free of arbitrage at any time $t \geq 0$, and if $S^\ast = \{S^\ast_t\} $ is a positive continuous semimartingale, then
    $X = \{X_t\}$ and $V = \{V_t\}$ are continuous semimartingales with
    \begin{equation*}
        \mathrm{d}V - X \, \mathrm{d}S^\ast  = - \frac{1}{2}f_\ast^{\prime\prime}(X)\, \mathrm{d}\langle X \rangle \leq 0.
    \end{equation*}
\end{thm}
The proof is given in Section~\ref{prthm2}.

\begin{rem}\upshape
Under the conditions of Theorem~\ref{thmn}, relative to the profit-and-loss by the model-free trading strategy $X$ in the external market
\begin{equation*}
    \hat{V}_T = V_0 + \int_0^T X_t \, \mathrm{d}S^\ast_t
\end{equation*}
with the same initial endowment $V_0$, the liquidity provider's wealth suffers from {\it Loss-Versus-Rebalancing} (''LVR'', cf.~\cite{MMRZ}), 
\begin{equation*}
   \hat{V}_T -  V_T  =  \frac{1}{2}\int_0^T f_\ast^{\prime\prime}(X_t)\, \mathrm{d}\langle X \rangle_t
\end{equation*}
that is positive by the convexity of the implicit function $f$ and non-degeneracy of the quadratic variation of $X = v^\prime(S^\ast)$ due to $\tau = 0$.
This extends~\cite{MMRZ,FMW1,CDM2} to cases with concentrated liquidity.
The use of Tanaka's formula instead of It\^o's formula is key to deal with concentrated liquidity
in the proof of Theorem~\ref{thmn}.
\end{rem}

\section{Dynamics under nonzero fees}\label{sec:reserves}
In this section, we consider the case $\tau > 0$.
We regard the LP reserves $X = \{X_t\}$ and $Y = \{Y_t\}$ as right-continuous stochastic processes with left-limits (RCLL).
Their left continuous modifications $X_- = \{X_{t-}\}$ and $Y_- = \{Y_{t-}\}$
are defined as $X_{t-} = \lim_{s\uparrow t}X_s$
and
$Y_{t-} = \lim_{s\uparrow t}Y_s$.
By definition, $\Delta X = X - X_-$ and $\Delta Y = Y-Y_-$.
Recall that $(X^f_t,Y^f_t)$ denotes the cumulative fee earning at time $t$. We define $X^f_-$, $Y^f_-$, $\Delta X^f$ and $\Delta Y^f$ similarly.

Extending Lemma~\ref{lem1} to the case $\tau>0$,
we define $S = s(X)$. 
The ask and bid price processes are then
\begin{equation*}
    A = \frac{1}{1-\tau} S, \ \ B = (1-\tau)S.
\end{equation*}
We define $S_-$, $A_-$, $B_-$, $\Delta S$, $\Delta A$, and
$\Delta B$ similarly.
Note that the relation
\eqref{eqS} in the case of $\tau=0$
is not extended to the case of $\tau>0$.

It is natural to assume that there are two types of traders; liquidity takers and arbitrageurs.
Let $X^o$ and $X^a$, respectively, denote the
cumulative flows of the first currency 
to the LP by liquidity takers and arbitrageurs.
We have $X = X_0 + X^o + X^a$ by definition.
Here, we assume $X^o_0 = X^a_0 = 0$ and
$B_0 \leq S^\ast_0 \leq A_0$.
The reserve of the second currency is determined as $Y_t = f_\ast(X_t)$ by the AMM algorithm.
The mathematical well-definedness of arbitrage-free reserve processes is ensured by the following theorem.
\begin{thm}\label{thm:main}
For any piecewise constant RCLL process $X^o$ with $X^o_0 = 0$ and a positive RCLL process $S^\ast$,
    there exists an RCLL process of finite variation $X^a$
    with $X^a_0 = 0$
     such that the LP is free of arbitrage at any time $t \geq 0$.
\end{thm}
The proof is given in Section~\ref{proof}.
\\

\noindent
For any swap $(\xi,\eta)$,
the corresponding fee earning is
\begin{equation*}
   ( \tau H(\xi)\xi, \tau H(\eta)\eta)
   = \frac{\tau}{1-\tau} ((\Delta x)^+, (\Delta y)^+)
\end{equation*}
for
\begin{equation*}
    (\Delta x, \Delta y) = ((1-\tau H(\xi))\xi, (1-\tau H(\eta))\eta).
\end{equation*}
We naturally assume that the reserve processes $X$ and $Y$ are of finite variation under $\tau > 0$.
Let 
\begin{equation}\label{dec}
    X = X^\uparrow - X^\downarrow, \ \ Y = Y^\uparrow - Y^\downarrow,
\end{equation}
where $X^\uparrow$, $X^\downarrow$,
$Y^\uparrow$ and $Y^\downarrow$ are nondecreasing processes. Then,
by the subadditivity of $x \mapsto (x)^+$,
we have
\begin{equation}\label{fee}
    \mathrm{d}X^f \geq \frac{\tau}{1-\tau} \mathrm{d}X^\uparrow, \ \ 
    \mathrm{d}Y^f \geq \frac{\tau}{1-\tau} \mathrm{d}Y^\uparrow.
\end{equation}
The equalities hold if there is no simultaneous orders more than two executed at any single moment.
\begin{thm}\label{thm:main2}
    If the LP is free of arbitrage at any time $t\geq 0$, and if $S^\ast$ is positive and continuous, then
    \begin{equation*}
        V -  \int_0^\cdot (X_t + X^f_t)\, \mathrm{d}S^\ast_t 
    \end{equation*}
    is nondecreasing under~\eqref{dec} and \eqref{fee}, where
    \begin{equation}\label{defint}
        \int_0^\cdot (X_t + X^f_t)\, \mathrm{d}S^\ast_t
        = (X + X^f)S^\ast - X_0S^\ast_0 - 
        \int_0^\cdot S^\ast_t \, \mathrm{d}X_t
         - \int_0^\cdot S^\ast_t \, \mathrm{d}X^f_t
    \end{equation}
    and the integrals in the right hand side are Stieltjes integrals.
\end{thm}
The proof is given in Section~\ref{prmain2}.

\begin{rem}\upshape
The definition \eqref{defint} follows 
a well-known approach to define a path integral with integrands of finite variation.
It is easy to see that the integral is the limit of natural Riemann approximations and so,
interpreted as the profit-and-loss for the trading strategy $X + X^f$ in the external market.
This is a model-free rebalancing strategy and 
provides a super-hedge of the impermanent loss: 
\begin{equation*}
   Y_0 +X_0 S^\ast_T -V_T  \leq  Y_0 +X_0 S^\ast_T 
   -V_0 - \int_0^T (X_t + X^f_t) \, \mathrm{d}S^\ast_t
   = \int_0^T (X_0-X_t - X^f_t) \, \mathrm{d}S^\ast_t.
\end{equation*}
by Theorem~\ref{thm:main2}.
We can extend the definition of the Loss-Versus-Rebalancing (LVR) by 
\begin{equation*}
     Y_0 +X_0 S^\ast_T -V_T - 
    \int_0^T (X_0-X_t - X^f_t) \, \mathrm{d}S^\ast_t.
\end{equation*}
Then, by the above inequality, the LVR is nonpositive when $\tau>0$ irrespectively of its size.
This fully extends the analysis in \cite{FMW2} for Uniswap v2 type architecture to Uniswap v3-type architecture.
We refer the reader to \cite{FMW2} for more discussions on the implication of the result.
\end{rem}

\begin{rem}\upshape
    If $S^\ast$ has jumps, the integration-by-parts formula gives
    \begin{equation*}
       \mathrm{d}V - (X_- + X^f_-) \mathrm{d} S^\ast_-  =  
    \mathrm{d}Y + \mathrm{d}Y^f +  S^\ast_- \mathrm{d}X +
    S^\ast_-\mathrm{d}X^f + \mathrm{d}[S^\ast, X+X^f],
    \end{equation*}
    where
    \begin{equation*}
        [S^\ast, X+X^f]_t = \sum_{u \leq t} \Delta S^\ast_u (\Delta X_u + \Delta X^f_u).
    \end{equation*}
Notice that if $\Delta S^\ast_u > 0$ induces an arbitrage trade, then $\Delta X_u < 0$ and $\Delta X^f = 0$.
 If $\Delta S^\ast_u < 0$ induces an arbitrage trade, then $\Delta X_u > 0$ and $\Delta X^f > 0$.
 Therefore the quadratic covariation term is negative and causes a positive LVR, which is a profit for arbitrageurs.
\end{rem}

\begin{rem}\upshape
A continuous-time model with continuous trajectories is naturally an approximation to the reality that is discrete by nature.
The faster the block production time in a blockchain, the greater the expected accuracy of the continuous model. 
Consistently with our result, it has been empirically observed by \cite{FC} that arbitrage losses for liquidity providers decrease when block production is faster.
An implication is that a speed-up of the block production will incentivise liquidity provision by mitigating the LVR. 
\end{rem}

\section{Multiple liquidity providers}
\subsection{Inf-convolution}
Here we consider an LP with multiple liquidity providers.
Suppose there are $N$ liquidity providers and each adopts a utility function $u^{(i)}$ and constants $\tau^{(i)}\geq 0$, $x_\ast^{(i)}\geq 0$ and $y_\ast^{(i)} \geq 0$ to create  what we call a subpool, $i=1,\dots,N$. 
Suppose that each initial reserves $(X_0^{(i)},Y^{(i)}_0)$ are chosen so that
the subpools are free of arbitrage at time $0$, that is,
\begin{equation*}
 -(1-\tau^{(i)}) f_\ast^{(i)\prime}(X_0^{(i)}) \leq 
    - \frac{1}{1-\tau^{(j)}} f_\ast^{(j)\prime}(X_0^{(j)})
\end{equation*}
for any $i$ and $j$ with $Y_0^{(i)}>0$ and $X_0^{(j)}>0$, where $f_\ast^{(i)}(x) = f^{(i)}(x + x_\ast^{(i)})$ and $f^{(i)}$ is the implicit function of $u^{(i)}$:
\begin{equation*}
 u^{(i)}(x, f^{(i)}(x)) = 
 u^{(i)}(X_0^{(i)} + x_\ast^{(i)}, Y_0^{(i)} + y_\ast^{(i)}), \ \ x >0.
\end{equation*}
The domain of $f_\ast^{(i)}$ is
\begin{equation*}
    \Xi^{(i)} := (-x_\ast^{(i)},\infty) \cap [0,x_\dagger^{(i)}], \ \ 
    x_\dagger^{(i)} = f_\ast^{(i)-1}(0+).
\end{equation*}
Let
\begin{equation*}
    \eta^{(i)}(\xi;x) = \frac{f_\ast^{(i)}(x + (1-\tau^{(i)} H(\xi))\xi) - f_\ast^{(i)}(x)}{1-\tau^{(i)} H(-\xi)}
\end{equation*}
for $ x \in \Xi^{(i)}$ and
such $\xi$ that $x + (1-\tau^{(i)} H(\xi))\xi \in  \Xi^{(i)}$.
Otherwise, we set
$\eta^{(i)}(\xi;x)= \infty$. 
This is the negative of the amount of the second currency to swap $\xi$ units of the first asset at the $i$th subpool when its reserve of the first currency is $x$.
Let 
\begin{equation}\label{infconv}
    \eta(\xi;x^{(1)},\dots,x^{(N)}) = \min\left\{
\sum_{i=1}^N \eta^{(i)}(\xi^{(i)},x^{(i)}); \sum_{i=1}^N \xi^{(i)} = \xi
    \right\}.
\end{equation}
This describes the negative of the total amount of the second currency 
a liquidity taker receives when she optimally allocates  $\xi$ units of the first currency among the subpools, where
$x^{(i)}$ is the reserve of the first currency at the $i$th subpool.
We remark that this is the inf-convolution of convex functions which has been studied in the context of optimal risk allocation; see, e.g.,~\cite{BE}.

Let $(X^{(i)},Y^{(i)})$ denote the reserve process for the $i$th subpool.
Consider 
an LP consisting of these $N$ subpools which
accepts a swap order $(\xi,\eta)$ at time $t>0$ if and only if 
\begin{equation*}
    \eta = \eta(\xi; X^{(1)}_{t-},\dots,X^{(N)}_{t-}).
\end{equation*}
If $(\xi_\ast^{(1)},\dots, \xi_\ast^{(N)})$  is the minimizer, that is,
\begin{equation*}
    \eta(\xi; X^{(1)}_{t-},\dots,X^{(N)}_{t-})
    = \sum_{i=1}^N \eta^{(i)}(\xi_\ast^{(i)},X^{(i)}_{t-}), \ \ 
    \sum_{i=1}^N\xi_\ast^{(i)}  = \xi,
\end{equation*}
the $i$th subpool's reserves are updated from $(X^{(i)}_{t-},Y^{(i)}_{t-})$ to
$$
(X^{(i)}_t,Y^{(i)}_t) =  (X^{(i)}_{t-} +  (1-\tau^{(i)} H(\xi_\ast^{(i)}))\xi_\ast^{(i)},  Y^{(i)}_{t-} + (1-\tau^{(i)} H(-\xi_\ast^{(i)})) \eta^{(i)}(\xi_\ast^{(i)},X^{(i)}_{t-})),
$$
and the $i$th cumulative fee accounts are updated
from $(X^{(i)f}_{t-},Y^{(i)f}_{t-})$ to
$$
(X^{(i)f}_{t},Y^{(i)f}_{t}) = 
 (X^{(i)f}_{t-} +  \tau^{(i)} H(\xi_\ast^{(i)})\xi_\ast^{(i)},  Y^{(i)f}_{t-} + \tau^{(i)} H(-\xi_\ast^{(i)}) \eta^{(i)}(\xi_\ast^{(i)},X^{(i)}_{t-})).
$$
The analysis in the preceding sections remains valid for each subpool. In other words, in order to study the wealth dynamics of a liquidity provider, 
it is not necessary to distinguish whether the liquidity provision is by creating an own LP or by joining the existing LP that adopts the inf-convolution algorithm.

\subsection{Unique internal price}\label{subsec:uip}
Uniswap v3 offers LPs with different fee tiers\footnote{At the time of this writing, there are 1\%, 0.3\%, 0.05\%, and 0.01\% fee tiers, cf.~\href{https://support.uniswap.org/hc/en-us/articles/20904283758349-What-are-fee-tiers}{Uniswap v3}}, each of which accepts concentrated liquidity provisions with different liquidity ranges.  
This means that an LP in Uniswap v3 consists of multiple liquidity providers with different $u_\ast$ but the same $\tau>0$.
Therefore, 
let us focus on the case that the transaction fee size is common: $\tau^{(i)} = \tau \in [0,1)$ for all $i$.
Fix $t>0$ and consider the following properties for the reserves $(X^{(i)}_{t-},Y^{(i)}_{t-})$:
\begin{enumerate}
    \item The internal price, $-f^{(j)\prime}(X^{(j)}_{t-})$, is common for all $j \in J_t$, where
\begin{equation*}
 J_t = \{j \in \{1,\dots,N\}; 
 X^{(j)}_{t-} \in \mathrm{int}(\Xi^{(j)})\},
\end{equation*}
where $ \mathrm{int}(\Xi^{(j)})$ is the set of 
the interior points of  $ \Xi^{(j)}$.
Let $S_{t-}$ denote the common internal price  $-f^{(j)\prime}(X^{(j)}_{t-})$.
\item $-f^{(i)\prime}(X^{(i)}_{t-}) \leq S_{t-}$
for any $i$ with $X^{(i)}_{t-}= 0$.
\item 
$-f^{(i)\prime}(X^{(i)}_{t-}) \geq S_{t-}$
for any $i$ with $Y^{(i)}_{t-} = 0$.
\end{enumerate}
Uniswap v3 indeed achieves these properties
by requiring a liquidity addition not to disturb the common internal price.
These properties are stable in time;
for brevity, let us consider the case that
\begin{enumerate}
    \item[2'.] $-f^{(i)\prime}(X^{(i)}_{t-}) < S_{t-}$
for any $i$ with $X^{(i)}_{t-} = 0$, and that
\item[3'.] $-f^{(i)\prime}(X^{(i)}_{t-}) >  S_{t-}$
for any $i$ with $Y^{(i)}_{t-}= 0$.
\end{enumerate}
Solving the minimization problem \eqref{infconv} using the Lagrange multiplier when $|\xi|>0$ is sufficiently small, we have
\begin{equation*}
   \sum_{j \in J_t}\xi_\ast^{(j)}  = \xi, \ \ 
    -\frac{1-\tau H(\xi)}{1 - \tau H(-\xi)}f^{(j)\prime}(X^{(j)}_{t-} + (1-\tau H(\xi)\xi_\ast^{(j)}) = \lambda
\end{equation*}
for all $j\in J_t$ for some $\lambda >0$.
Therefore,
the updated internal price 
$$-f^{(j)\prime}(X^{(j)}_t)
= -f^{(j)\prime}(X^{(j)}_{t-} + (1-\tau H(\xi)\xi_\ast^{(j)})
$$
is again common for all $j \in J_t$. 

\subsection{Constant Product Market}
As an example, consider the case of a Constant Product Market Maker with Concentrated Liquidity with a common transaction fee size, that is, 
$u^{(i)}(x,y) = xy$ and $\tau^{(i)} = \tau \in [0,1)$ for all $i = 1,\dots,N$.
We will show that the subpool model~\eqref{infconv} describes the architecture of Uniswap v3.
First, note that
\begin{equation*}
    f_\ast^{(j)}(x) = \frac{k_j^2}{x + x_\ast^{(j)}} - y_\ast^{(j)},  \ \ k_j =  \sqrt{
    (X_0^{(j)}  + x_\ast^{(j)}) (Y_0^{(j)}  + y_\ast^{(j)})}
\end{equation*}
and so,
\begin{equation*}
    \eta^{(j)}(\xi;X^{(j)}_{t-}) =  
    \frac{1}{1-\tau H(-\xi)} \left(
    \frac{k_j^2}{(1-\tau H(\xi))\xi + X^{(j)}_{t-} + x_\ast^{(j)}} - Y^{(j)}_{t-}-y_\ast^{(j)} \right).
\end{equation*}
Fix $t > 0$ and suppose that the internal prices
\begin{equation*}
 -f_\ast^{(j)\prime}(X^{(j)}_{t-}) = \frac{Y^{(j)}_{t-} + y_\ast^{(j)}}{X^{(j)}_{t-}+ x_\ast^{(j)}}
\end{equation*}
satisfy the properties 1, 2' and 3' in Section~\ref{subsec:uip}.
Solving the minimization problem \eqref{infconv} using the Lagrange multiplier when $|\xi|$ is sufficiently small, we have
\begin{equation*}
   \sum_{j \in J_t}\xi_\ast^{(j)}  = \xi, \ \ 
     \sqrt{\frac{1-\tau H(\xi)}{1 - \tau H(-\xi)}}
    \frac{k_j}{(1-\tau H(\xi))\xi^{(j)}_\ast + X^{(j)}_{t-} + x_\ast^{(j)}} = \sqrt{\lambda}
\end{equation*}
for all $j\in J_t$ for some $\lambda >0$.
Putting
\begin{equation*}
    k = \sum_{j \in J_t} k_j, \ \ X_t = \sum_{j \in J_t} X^{(j)}_t, \ \ 
    x_\ast = \sum_{j \in J_t} x_\ast^{(j)}, \ \ 
    \ \ Y_t = \sum_{j \in J_t} Y^{(j)}_t, \ \ 
      y_\ast = \sum_{j \in J_t} y_\ast^{(j)},
\end{equation*}
we have
\begin{equation*}
   \sqrt{\frac{1-\tau H(\xi)}{1 - \tau H(-\xi)}} \frac{k}{\sqrt{\lambda}} = (1-\tau H(\xi))\xi + X_{t-} + x_\ast
\end{equation*}
and so,
\begin{equation*}
\begin{split}
     \eta(\xi,X^{(1)}_{t-},\dots,X^{(N)}_{t-}) & =\sum_{j \in J_t}
     \frac{1}{1-\tau H(-\xi)} \left(
     \frac{k_j^2}{(1-\tau H(\xi))\xi^{(j)}_\ast + X^{(j)}_{t-} + x_\ast^{(j)}} - Y^{(j)}_{t-} - y^{(j)}_\ast \right)\\
     &=  \frac{1}{1-\tau H(-\xi)} \left(
     \sqrt{\lambda \frac{1-\tau H(-\xi)}{1-\tau H(\xi)}}k
     - 
     Y_{t-} - y_\ast \right)
     \\
    & = 
     \frac{1}{1-\tau H(-\xi)} \left(
    \frac{k^2}{(1-\tau H(\xi))\xi + X_{t-} + x_\ast} - Y_{t-} - y_\ast \right).
\end{split}
\end{equation*}
Thus, the liquidity provision is as if there is a single Constant Product Market Maker with Concentrated Liquidity
with reserves $(X_t,Y_t)$:
\begin{equation*}
    ( X_t  + x_\ast)( Y_t + y_\ast) = k^2.
\end{equation*}
The reserve processes of subpools satisfy
\begin{equation*}
    \frac{k_j}{X_t^{(j)} + x_\ast^{(j)}} =
    \sqrt{\lambda \frac{1-\tau H(-\xi)}{1-\tau H(\xi)}} = \frac{k}{X_t + x_\ast}, \ \ j \in J_t
\end{equation*}
and so,
\begin{equation*}
    \mathrm{d}X_t^{(j)} = \frac{k_j}{k}\mathrm{d}X_t, \ \ 
    \mathrm{d}Y_t^{(j)} = \frac{k_j}{k}\mathrm{d}Y_t.
\end{equation*}
This further implies that the fee income for the $j$th subpool is the $k_j/k$ portion of the total fee income.
In other words, a fee income of the total pool is allocated to its subpools according to their portions.
The updated internal price
\begin{equation*}
    \frac{Y^{(j)}_{t} + y_\ast^{(j)}}{X^{(j)}_t+ x_\ast^{(j)}}
    = \frac{k_j^2}{(X^{(j)}_t+ x_\ast^{(j)})^2}
    = \frac{k^2}{(X_t + x_\ast)^2} = \frac{Y_t + y_\ast}{X_t + x_\ast}
\end{equation*}
remains common for all $j \in J_t$.
These properties are consistent with the description of Uniswap v3 in Section~3 of \cite{CDM2}.
\begin{rem}\upshape
    The considerations in this section demonstrate an important feature of Constant Function Market Makers with concentrated liquidity and with a common fee tier (such as in Uniswap v3): 
    Their construction is optimal in the sense that the distribution of fee income proportional to the liquidity depth of each liquidity provider follows naturally from the optimal allocation~\eqref{infconv} of the amount $\xi$ sent to the LP's subpools. 
\end{rem}

\section{Conclusion}
We have given a mathematical formulation of Automated Market Makers in Decentralized Exchanges, which in particular includes the popular platform Uniswap v3 as an example.
The preceding studies on Impermanent Loss and Loss-Versus-Rebalancing have been extended to this rigorous framework incorporating concentrated liquidity provision and transaction fee collection.
In particular, we have shown that Impermanent Loss can be super-hedged by a model-free rebalancing strategy in the external market if the external market price process is continuous. 
We have ascertained that the Uniswap v3-type architecture for multiple liquidity providers can be described as a liquidity pool which optimally allocates a swap order to its subpools, each of which is a liquidity pool created by each liquidity provider.
Hence, the dynamics of a liquidity provider's wealth does not depend on whether she creates a new pool or joins an existing pool.\\

\noindent
{\bf Statements and Declarations:}
The authors have no conflicts of interest to declare.


\begin{appendix}
\section{Proofs}
\subsection{Proof of Lemma~\ref{lem1}}
\label{prlem1}
Note that $X_t = 0$ can happen only if $x_\ast > 0$ since $f(0+) = \infty$.
In this case, $0 \in \Xi$ and $r = s(0)$.
If $X_t =0$, then $Y_t = f_\ast(X_t) > 0$ and $r = s(0) = s(X_t) = B_t \leq S^\ast_t$ by \eqref{NAL}.
Therefore, we have $S_t = r = s(X_t)$.
Similarly,
note that $Y_t = 0$ can happen only if $y_\ast > 0$ since $f(\infty-) = 0$. In this case, $x_\dagger \in \Xi$ and
$l = s(x_\dagger)$.
If $Y_t = 0$, then $X_t = f_\ast^{-1}(0) = x_\dagger$ and
$l = s(x_\dagger) = s(X_t) = A_t \geq S^\ast_t$ by \eqref{NAR}.
Therefore, we have $S_t = l = s(X_t)$.
When $X_t > 0$ and $Y_t > 0$, we have both \eqref{NAR} and \eqref{NAL}, so $s(X_t) =S^\ast_t$.
In particular, $S^\ast_t \in s(\Xi)$ and so,
$S_t = S^\ast_t = s(X_t)$.

\subsection{Proof of Theorem~\ref{thm1}}
\label{prthm1}
Recall that $f_\ast$ is strictly convex, decreasing and continuously differentiable on $\Xi$.
When $\Xi$ is not an open set, the last property means that that $f_\ast$ can be extended to a continuously differentiable function on an open set $O$ with
$\Xi \subset O$.
Indeed, we can take $O:=(-x_\ast,\infty)$.
If $p \in s(\Xi)$, we have
$x^\ast := (f_\ast^\prime)^{-1}(-p)=s^{-1}(p) \in \Xi$ and 
\begin{equation*}
    v(p) = x^\ast p + f_\ast( x^\ast), \ \ v^\prime(p) = x^\ast
\end{equation*}
since $f_\ast^\prime(x^\ast) = -p$.
On the other hand,
since the LP is free of arbitrage with
$S^\ast_t \in s(\Xi)$, by Lemma~\ref{lem1},
we have $X_t = s^{-1}(S^\ast_t) = f_\ast^{-1}(-S^\ast_t)$
and so,
\begin{equation*}
    V_t = X_tS^\ast_t + Y_t = 
    X_t S^\ast_t +  f_\ast(X_t)  = v(S^\ast_t)
\end{equation*}
and $v^\prime(S^\ast_t) = X_t$ as claimed.

\subsection{Proof of Theorem~\ref{thmn}}
\label{prthm2}
By Tanaka's formula (Theorem (1.2) and Proposition (1.3) of Chapter VI, Revuz and Yor~\cite{RY}),
$S = \{S_t\}$ is a continuous semimartingale with
\begin{equation*}
    \mathrm{d}S = 1_{(l,r)}(S^\ast)\, \mathrm{d}S^\ast + 1_{\{l\}}(S^\ast)\, \mathrm{d}\Lambda^l - 1_{\{r\}}(S^\ast)\, \mathrm{d}\Lambda^r,
\end{equation*}
where $\Lambda^l$ and $\Lambda^r$ are nondecreasing processes.
This implies in particular that
\begin{equation}\label{sandsast}
    S \,\mathrm{d}S = S^\ast\, \mathrm{d}S, \ \ 
     \mathrm{d}\langle S^\ast , S \rangle =  \mathrm{d}\langle S \rangle, \ \ 
     S\, \mathrm{d}\langle S \rangle
     = S^\ast\,\mathrm{d}\langle S \rangle.
\end{equation}
On the other hand,
by Lemma~\ref{lem1}, we have $X = s^{-1}(S)$
and so, $X$ is a $\Xi$-valued continuous semimartingale by It\^o's formula.
By \eqref{sandsast}, we have
\begin{equation*}
    S^\ast\, \mathrm{d}X = S \, \mathrm{d}X, \ \ 
    \mathrm{d}\langle S^\ast , X \rangle =  \mathrm{d}\langle S , X \rangle.
\end{equation*}
Recalling $S = s(X) = -f_\ast^\prime(X)$ and $Y = f_\ast(X)$, we have then by It\^o's formula,
\begin{equation*}
\begin{split}
     \mathrm{d}V &= X\, \mathrm{d}S^\ast  + 
    S \, \mathrm{d}X + \mathrm{d}\langle S, X \rangle + 
    \mathrm{d}Y
    \\
    & =  X \, \mathrm{d}S^\ast  
    -f_\ast^\prime(X) \, \mathrm{d}X - f_\ast^{\prime\prime} (X)\, \mathrm{d}\langle  X \rangle + 
    \mathrm{d}Y
    \\
    &=  X \, \mathrm{d}S^\ast - \frac{1}{2}f_\ast^{\prime\prime}(X)\, \mathrm{d}\langle X \rangle
\end{split}
\end{equation*}
as claimed.

\subsection{Proof of Theorem~\ref{thm:main}}\label{proof}
First we consider the case $x_\ast = y_\ast = 0$ and hence
$\Xi = (0,\infty)$ (see Remark~\ref{remXi}).
Let $\psi = \log S^\ast - \log S_0- \log (1-\tau)$
and $a = -2 \log (1-\tau)$.
Note that $0 \leq \psi_0 \leq a$ by $B_0 \leq S^\ast_0 \leq A_0$.
Let
$(\phi,\eta)$ be the solution of the Skorokhod problem on $[0,a]$ for the path $\psi$.
See \cite{KLRS} for an explicit solution.
By definition, 
$\eta$ is of finite variation and 
\begin{equation*}
0 \leq \phi = \psi + \eta \leq a.
\end{equation*}
Let $X = s^{-1}(S_0\exp(-\eta))$ so that 
$\log S - \log S_0 = - \eta$. Then,
\begin{equation*}
        \mathrm{d} \phi = \mathrm{d}\psi + \mathrm{d}\eta = \mathrm{d}  \log \frac{S^\ast}{(1-\tau) S} .
\end{equation*}
Since $\phi_0 = \psi_0=
\log S^\ast_0 - \log S_0- \log (1-\tau)$, we then conclude
\begin{equation*}
\phi = \log \frac{S^\ast}{(1-\tau) S}.
\end{equation*}
Note that $0 \leq \phi \leq a$ is equivalent to 
$B \leq S^\ast \leq A$.
We can take $X^a = X - X_0 - X^o$.

In the general case, since $X^o$ is piecewise constant by assumption,
the time interval can be separated into intervals in which $\Delta X^o = 0$.
At each time $t$ with $\Delta X^o_t \neq 0$, 
it is trivally possible to take $\Delta X^a_t$ such that
the LP is free of arbitrage at time $t$.
Therefore, we assume without loss of generality that $X^o = 0$. Let $X$ be as above and 
\begin{equation*}
    \sigma = \inf\{t \geq 0; X_t < 0 \text{ or } Y_t < 0\}.
\end{equation*}
Then, the process $X^a := X -X_0$ meets the requirement up to $\sigma$.
In the case $X_\sigma < 0$,
we modify $X$ as $X_t = 0$ for $\sigma \leq t < \sigma^\prime$,
where
\begin{equation*}
    \sigma^\prime = \inf\{t \geq \sigma; S^\ast_t < b(0)\}.
\end{equation*}
In the case $Y_\sigma < 0$,
we modify $X$ as $X_t = x_\dagger$ for $\sigma \leq t < \sigma^\prime$,
where
\begin{equation*}
    \sigma^\prime = \inf\{t \geq \sigma; S^\ast_t > a(x_\dagger)\}.
    \end{equation*}
    In any case, $\sigma^\prime > 0$ and 
    the LP is free of arbitrage up to $\sigma^\prime$. We can repeat the same argument to concatanate $X$.

\subsection{Proof of Theorem~\ref{thm:main2}}
\label{prmain2}
Since $Y = f_\ast(X)$, $S = -f_\ast^\prime(X)$ and $f$ is convex, we have
\begin{equation*}
    \mathrm{d}Y^\uparrow \geq  S_- \, \mathrm{d}X^\downarrow, \ \ 
      \mathrm{d}Y^\downarrow \leq  S_- \, \mathrm{d}X^\uparrow.
\end{equation*}
Therefore, if
$S^\ast$ is continuous, then
\begin{equation*}
\begin{split}
       \mathrm{d}V - (X_- + X^f_-) \, \mathrm{d} S^\ast  &=  
    \mathrm{d}Y + \mathrm{d}Y^f +  S^\ast \, \mathrm{d}X +
    S^\ast\, \mathrm{d}X^f\\
    &\geq \left(1 + \frac{\tau}{1-\tau} - \frac{S^\ast}{S_-}\right)\,  \mathrm{d}Y^\uparrow + 
    \left(S^\ast + \frac{\tau}{1-\tau}S^\ast - S_-\right)\, \mathrm{d}X^\uparrow\\
    &= \frac{1}{S_-} (A_- - S^\ast) \, \mathrm{d}Y^\uparrow
    + \frac{1}{1-\tau}(S^\ast - B_-) \, \mathrm{d}X^\uparrow.
\end{split}
\end{equation*}
The right hand side is nonnegative 
since the LP is free of arbitrage.

\end{appendix}

\end{document}